\documentclass[%
reprint,
amsmath,amssymb,
aps, prl, groupedaddress,
longbibliography,
]{revtex4-2}

\usepackage{graphicx}
\usepackage{dcolumn}
\usepackage{bm}
\usepackage{natbib}

\newcommand{\ytterbium}{$^{171}$Yb$^+$}

\begin{document}
	
\preprint{APS/123-QED}

\title{Integrated optical addressing of a trapped ytterbium ion}

\author{M.~Ivory}
\email{mkivory@sandia.gov}
\affiliation{Sandia National Laboratories, Albuquerque, New Mexico 87185, USA}
\author{W.~J.~Setzer}
\affiliation{Sandia National Laboratories, Albuquerque, New Mexico 87185, USA}
\author{N.~Karl}
\affiliation{Sandia National Laboratories, Albuquerque, New Mexico 87185, USA}
\author{H.~McGuinness}
\affiliation{Sandia National Laboratories, Albuquerque, New Mexico 87185, USA}
\author{C.~DeRose}
\affiliation{Sandia National Laboratories, Albuquerque, New Mexico 87185, USA}
\author{M.~Blain}
\affiliation{Sandia National Laboratories, Albuquerque, New Mexico 87185, USA}
\author{D.~Stick}
\affiliation{Sandia National Laboratories, Albuquerque, New Mexico 87185, USA}
\author{M.~Gehl}
\affiliation{Sandia National Laboratories, Albuquerque, New Mexico 87185, USA}
\author{L.~P.~Parazzoli}
\affiliation{Sandia National Laboratories, Albuquerque, New Mexico 87185, USA}

\date{\today}

\begin{abstract}
We report on the characterization of heating rates and photo-induced electric charging on a microfabricated surface ion trap with integrated waveguides. Microfabricated surface ion traps have received considerable attention as a quantum information platform due to their scalability and manufacturability. Here we characterize the delivery of 435~nm light through waveguides and diffractive couplers to a single ytterbium ion in a compact trap. We measure an axial heating rate at room temperature of 0.78$\pm$0.05~q/ms and see no increase due to the presence of the waveguide. Furthermore, the electric field due to charging of the exposed dielectric outcoupler settles under normal operation after an initial shift. The frequency instability after settling is measured to be 0.9~kHz. 
\end{abstract}


\maketitle

\section{Introduction}
\label{sec:intro}
Though quantum information applications have motivated most of the development of microfabricated surface ion traps \cite{stick:2005, seidelin:2006}, many of the same advantages and challenges are also applicable to atomic clocks. This is particularly true for those using optical transitions, which require atomic confinement within the Lamb Dicke regime. While neutral atoms achieve this confinement in an optical lattice with many atoms, a conventional sized ion trap is limited to small numbers of crystallized ions which limit the signal-to-noise-ratio. Despite this limitation, an ion atomic clock using a single ion in a macro-scale trap has achieved unprecedented fractional systematic uncertainties better than $1\times10^{-18}$ using quantum logic spectroscopy, albeit in a laboratory scale environment \cite{brewer:2019a, brewer:2019b}. 

Furthermore, microfabricated ion traps have been extended to support many independently addressable individual trapping sites. This capability maintains high performance in a deployable configuration that combines low size, weight, and power (SWaP). Additional ions can also be used to replace lost ions, monitor environmental conditions, eliminate Dick effect errors using staggered optical interrogation sites, and optimize feed back to the local oscillator using different interrogation times to further extend the coherence time. This latter benefit fundamentally improves the clock stability scaling \cite{borregaard:2013}. Finally, the ability to integrate waveguides, detectors, and other photonic elements into the surface trap adds scalability and robustness compared to the alternative of free-space optics \cite{mehta:2016,jiang:2011,slichter:2017,mehta:2020}.

The benefits provided by microfabricated traps have led to significant advances, but several challenges remain, both in the microfabricated trap integration and in understanding the added systematic effects on clock operation due to the integrated elements. First, optical transitions of many trapped ion species lie in the blue to ultraviolet spectrum where transmission through standard silicon nitride waveguides is inefficient. Instead, one must use aluminum nitride or alumina to achieve low loss \cite{west:2019,sorace:2019}. To date, there has been one demonstration of trapped ion manipulation using waveguide delivery of blue light, which included the photoionization and Doppler cooling wavelengths for Sr$^+$ at 405~nm, 461~nm, and 422~nm \cite{niffenegger:2020}. 

A second outstanding challenge for microfabricated traps is anomalous heating, which couples noise to the motional state of the ion. Previous studies \cite{brownnutt:2015, boldin:2018, sedlacek:2018, hite:2012} have found that the heating rate dependence of a trapped ion scales with the ion-electrode distance approximately as d$^{-4}$. Furthermore, this heating rate is expected to increase considerably over a dielectric as opposed to a metallic layer \cite{kumph:2016}. This is a potential complication to photonics integration which is fabricated with dielectric materials. 

Finally, the presence of light on a dielectric material creates electrostatic fields through photo-induced charging. This charging impacts the performance and operation of the clock by moving the ion relative to the optical beams, increasing micromotion, and creating systematic shifts in the transition frequency. 

\begin{figure}[ht]
\resizebox{.48\textwidth}{!}{
\includegraphics{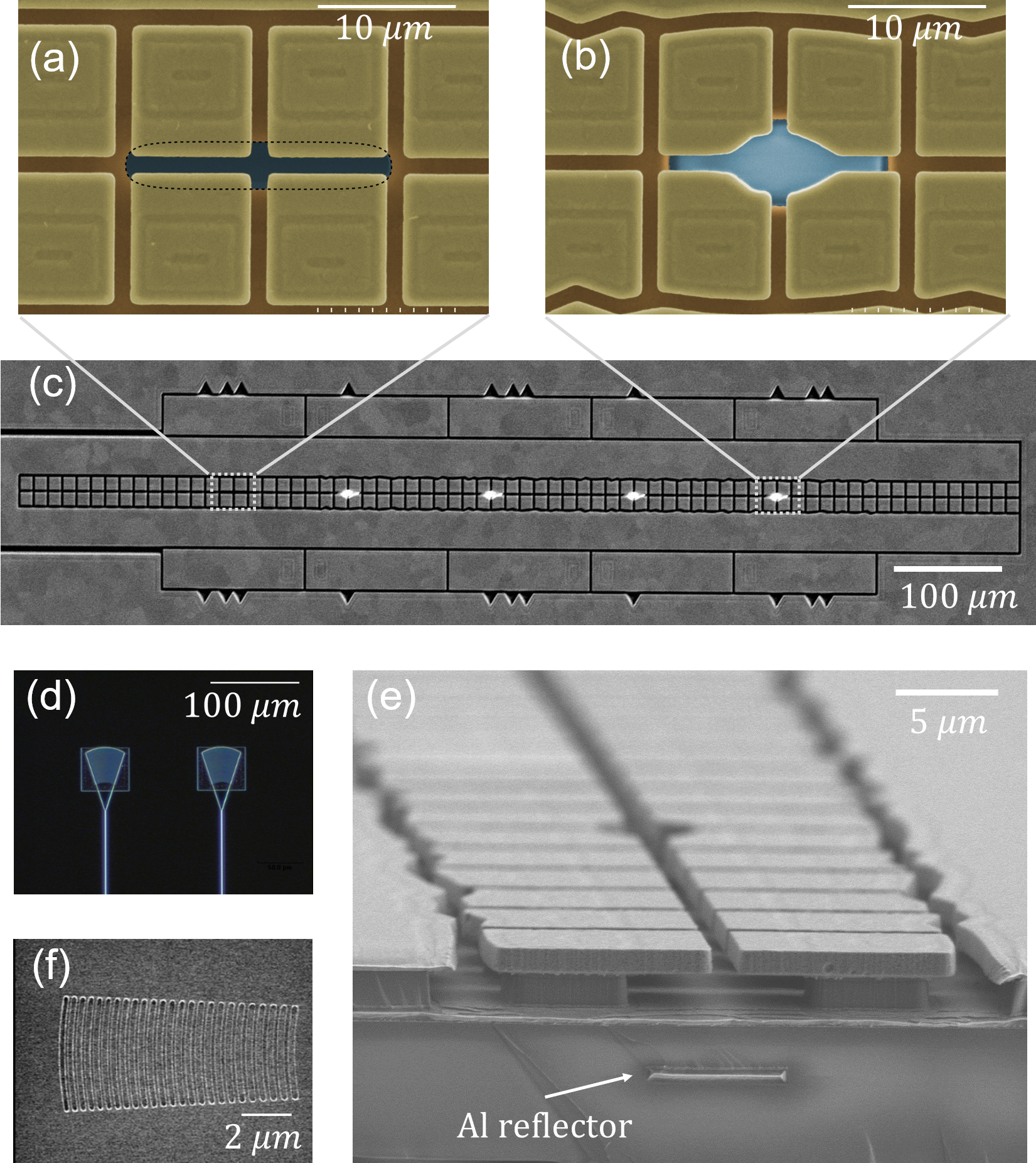}}
\caption{(a) False-colored scanning electron micrograph (SEM) image of the loading hole region of the trap and DC control electrodes in the topmost metal layer. Dark gray and dotted line indicates cutout beneath the overhanging metal layer to allow backside loading. (b) False-colored SEM image of the region above the leftmost output diffraction grating, shown in blue beneath the overhanging metal layer. (c) SEM image of the trapping region of the surface trap. Smallest electrodes vertically centered are DC control electrodes, with RF electrodes immediately above and below and DC bias electrodes furthest away. Triangles above and below DC bias electrodes are for beam alignment. The trap region shows the loading hole on the left and four output diffraction gratings. (d) Dark field optical micrograph image of the input diffraction gratings for coupling light to the waveguides. (e) SEM imaged sectional view of the trap showing the overhanging top metal electrodes and Al reflector below an output grating. (f) SEM image of a diffractive output coupler etched in SiN$_x$ for coupling light from the waveguides and focusing light 20 $\mu$m above the chip surface.}
\label{fig:trapWG}
\end{figure}

In this work we study the electric field and the heating rate impact on an ion in proximity to an integrated photonic out-coupler that is used to deliver light at 435~nm to a trapped $^{171}$Yb$^+$ ion 20~$\mu$m above the trap surface (Fig.~\ref{fig:trapWG}). This ion height is the lowest published to-date. Ion heights in surface traps are a concern because of the correlation of the heating rate of the ion to its height above the trap surface. Despite the increase in ion heating, lower heights are desirable, particularly in architectures in which each ion is confined in an individual trapping site or multiple trapping sites are required on a single chip, such as the QCCD architecture~\cite{brown:2016, pino:2021}. In these cases, the lower ion height allows for closer ion spacing, due to the smaller geometric size of the DC trapping electrodes, and thus more ions can be stored on a single chip. Additionally, the lower ion height also reduces the width of the RF electrode, which in turn reduces the overall capacitance and thus RF power dissipation of the trap chip---an advantage in SWaP critical applications such as a deployable optical atomic clock.

The trap includes integrated waveguides that connect diffractive grating input and output couplers below the top metal level; using the integrated photonics, we have demonstrated hyperfine spectroscopy, ground state motional sideband cooling, and heating rate measurements. We compare heating rates over a range of 80~$\mu$m along the trap to measure the impact of exposed dielectric on the ion. We measured heating rates of 0.78$\pm$0.05~q/ms, which agrees with previously reported data for larger ion-surface distances following a d$^{-4}$ scaling. Surprisingly, we see no increase in the heating rates as the ion approaches the dielectric grating. This gives encouraging evidence that non-CMOS compatible fabrication steps such as indium tin oxide (ITO) coatings may not be necessary to achieve adequate heating rates in photonics integrated chips. However, we do observe photo-induced charging of the dielectric which leads to a shift in the measured trap frequency.

\section{Trap design and fabrication}
\label{sec:design}

The trap used in this work was designed and fabricated at Sandia National Laboratories using CMOS compatible silicon MEMS fabrication technologies in a configuration similar to those previously reported \cite{moehring:2011}. Images of the trap are shown in Figure~\ref{fig:trapWG}. Multiple metal layers are used to route RF and DC signals to electrodes on the top metal level. The top metal layer, containing the RF routing and trap DC control electrodes, overhangs the underlying SiO$_2$ dielectric layer to eliminate direct line of sight between the ion and dielectric. Notable to this particular trap is the integration of silicon nitride waveguides embedded in the SiO$_2$ inter-metal dielectric and routed between metal levels and around vertical electrical vias made of chemical vapor deposited tungsten for delivery of 435~nm light. Light is coupled in and out of the waveguide using diffractive gratings. Figure~\ref{fig:trapWG}(c) shows the trapping region with a loading hole on the left, and four output gratings to its right. Figure~\ref{fig:trapWG}(a) shows a zoom-in of the loading hole site, and Figure~\ref{fig:trapWG}(b) shows a zoom-in of the leftmost output grating site. The small electrodes immediately above and below the trap axis are DC control electrodes. The RF electrodes are the long rails between the DC control electrodes and the outer DC bias electrodes.

Plasma-enhanced chemical vapor deposition (PECVD) silicon nitride waveguides with a thickness of 70~nm were deposited within the ~2 $\mu$m SiO$_2$ dielectric layer between trap metal layers; the silicon nitride layer is kept thin to reduce material absorption. We have observed a propagation loss of -2.5~dB/cm at 435~nm in test waveguides fabricated with similar processes. The input grating (Figure~\ref{fig:trapWG}(d)) is designed for free space coupling with a relatively large beam waist of 15~$\mu$m. The output grating (Figure~\ref{fig:trapWG}(f)) is located between the DC electrodes and designed to produce a beam which focuses at the height of the ion. Utilizing the metal trap routing layer, an Al reflector is positioned below the output grating to increase efficiency (Figure~\ref{fig:trapWG}(e)). 

Simulations of the ideal structure show -6.5~dB input coupling loss and -1.6~dB output coupling loss for an expected total through chip insertion loss of approximately -9.7~dB. The output beam is simulated to emit at a $63^\circ$ angle with respect to the plane of the chip and have a focused beam waist of 1.45~$\mu$m at the height of the ion with expected polarization parallel to the chip surface. In a separate die external to the ultra high vacuum chamber trapped ion system we measure a through chip insertion loss of -22~dB and a beam waist of 2.5~$\mu$m at a height of 20~$\mu$m above the chip. The difference in the beam waists is likely due to fabrication tolerances, while the difference in through chip insertion loss is likely due to reduced input coupling from a cleaved fiber. In future devices, we plan to use edge coupling instead of the input diffraction grating, which is expected to reduce input coupling loss to -3~dB with a photolithography process and to -0.5~dB with e-beam lithography.

\section{Experimental details and results}
\label{sec:experiment}

\ytterbium ions are produced via an isotope-selective two-photon ionization process \cite{johanning:2011}, Doppler cooled to an average number of quanta $\langle n \rangle \approx 3$ quanta, and sideband cooled to achieve $\langle n \rangle$ as low as 0.1 quanta. An RF voltage at 74.5~MHz is applied to the RF rails (yellow in Figure~\ref{fig:trapWG}) and quasi-static voltages are applied to electrodes to control the axial position and cancel stray fields. Typical secular axial trap frequencies are tunable between 2$\pi \times$2.5 and 2$\pi \times$5.2~MHz and radial trap frequencies are 2$\pi \times$12.7~MHz. These frequencies are verified via spectroscopy on the $|2S_{1/2},F=0\rangle$ to $|2D_{3/2},F=2\rangle$ quadrupole clock transition. All beams are delivered via free-space optics with the exception of the 435~nm beam, which addresses the quadrupole transition and can be switched between free-space and waveguide delivery.

\subsection{Heating rates}

We measure the motional heating rate $\langle \dot{n}\rangle$, or the time rate of change of the average number of quanta $n$ for a given motional mode, for the axial trap mode on the $|2S_{1/2},F=0,m_F=0\rangle$ to $|2D_{3/2},F=2,m_F=+2\rangle$ transition 
by comparing the amplitudes of the red and blue axial sidebands \cite{roos:2000}. The $\Delta m_F = +2$ transition was chosen because it has the strongest coupling with the freespace beam for our geometry.

%

%

\begin{figure}[ht]
	\resizebox{.48\textwidth}{!}{
		\includegraphics{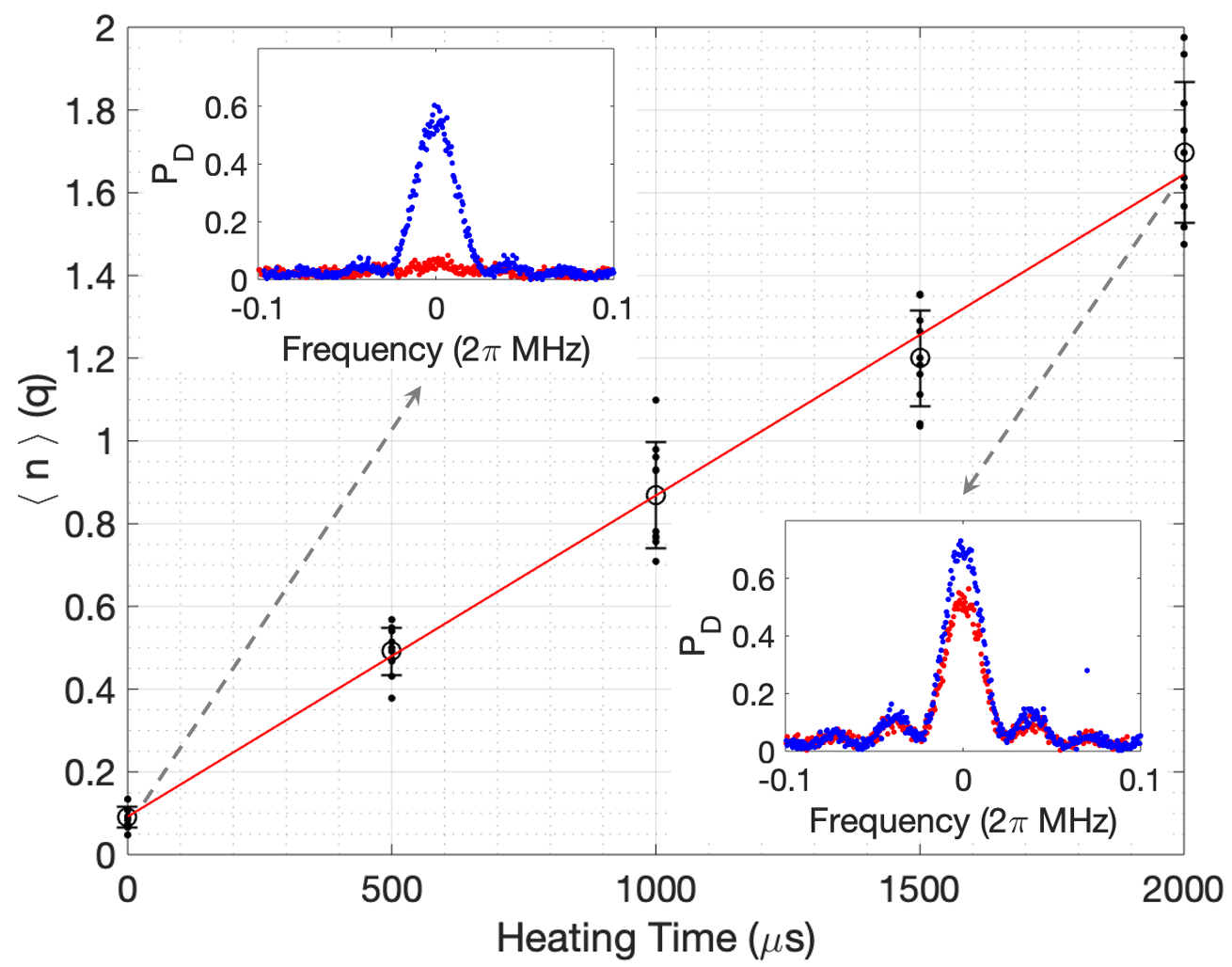}}
	\caption{$\langle n \rangle$ as a function of time using sideband amplitudes over the loading hole. Insets show the corresponding sideband amplitude measurements for $\langle n \rangle = 0.1$ (top) and $\langle n \rangle = 3$ (bottom). These measurements were taken using a freespace clock beam.The red and blue data corresponds to the red and blue sidebands.}
	\label{fig:heatingSideband}
\end{figure}

Results of a heating rate measurement taken with the sideband asymmetry method can be seen in Figure~\ref{fig:heatingSideband}. After sideband cooling and without any additional heating time, we achieve $\langle n \rangle \sim 0.1$ and the red axial sideband is significantly suppressed. After a heating time of 3~ms, the ion reaches a temperature roughly the same as the Doppler cooling temperature, or $\langle n \rangle \sim 3$ quanta, so we limit our heating time in a typical measurement to 2~ms or less, as shown in Figure~\ref{fig:heatingSideband}.

%

\begin{figure}[ht]
\resizebox{.45\textwidth}{!}{
\includegraphics{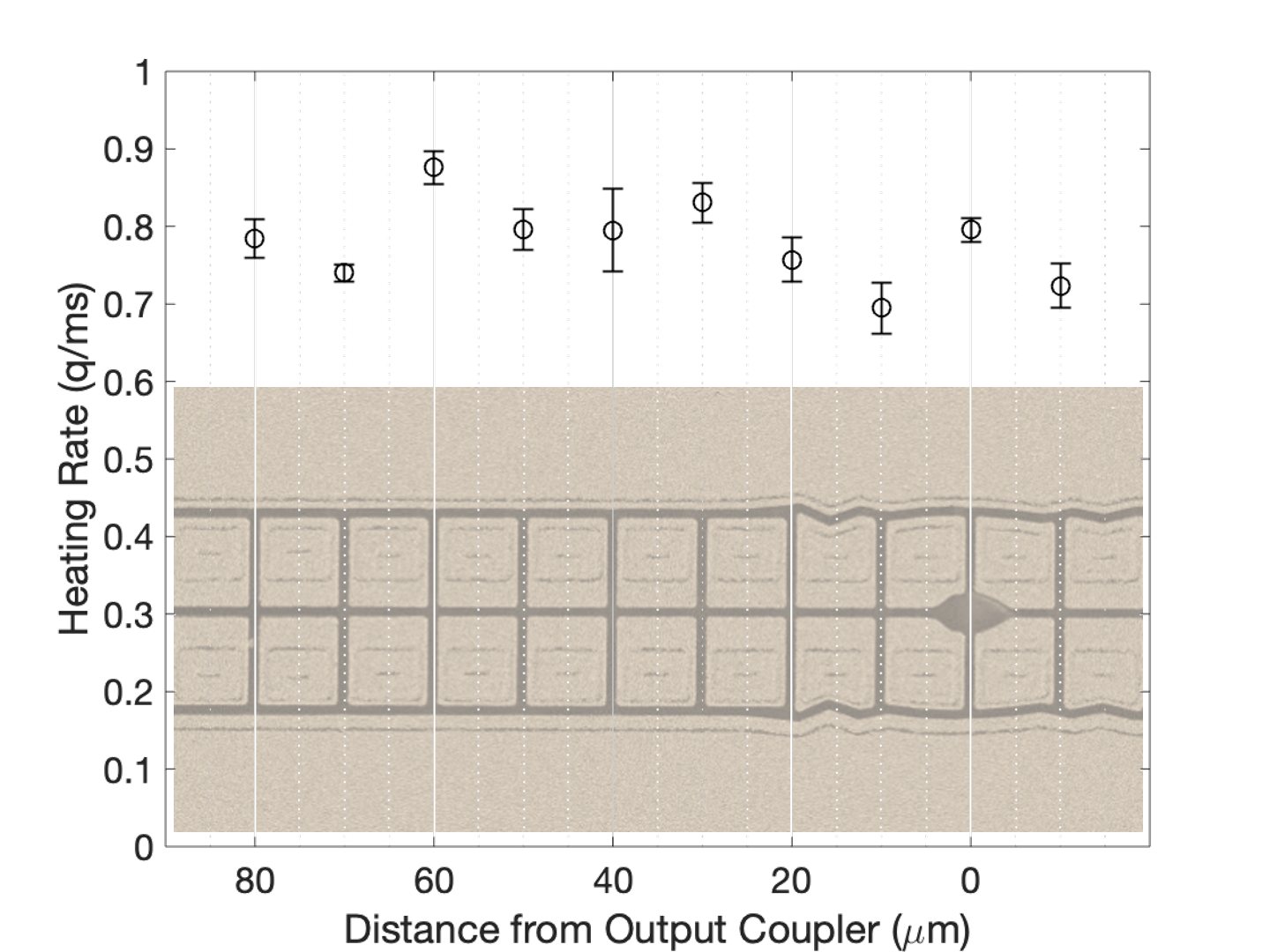}}
\caption{Heating rate as a function of position overlayed on the trap region, where the output coupler for the waveguide is at 0~$\mu$m, the loading hole is at 80~$\mu$m, and the maximum overlap between the ion and the waveguide output light is at 70~$\mu$m. No measurable change in heating rate is observed as the ion is shuttled from the loading hole to the output grating where it is closest to exposed dielectric material.}
\label{fig:HRvPosition}
\end{figure}

To address the concern of increased heating rates over dielectric materials \cite{kumph:2016}, we shuttle our ion from the loading hole, where it is above bare aluminum electrodes, to a location directly over the output grating of a waveguide. Here the ion is in direct line of sight with the silicon nitride waveguide. Figure~\ref{fig:HRvPosition} shows the heating rate at 10~$\mu$m increments over a range of 80~$\mu$m between the loading hole and the dielectric output grating. Remarkably, there is no measurable change in heating rate even when the ion is positioned directly above the dielectric output grating at a height of 20~$\mu$m. The mean heating rate across the chip is 0.78$\pm$0.05~q/ms. 

\begin{figure}[ht]
	\resizebox{.45\textwidth}{!}{
		\includegraphics{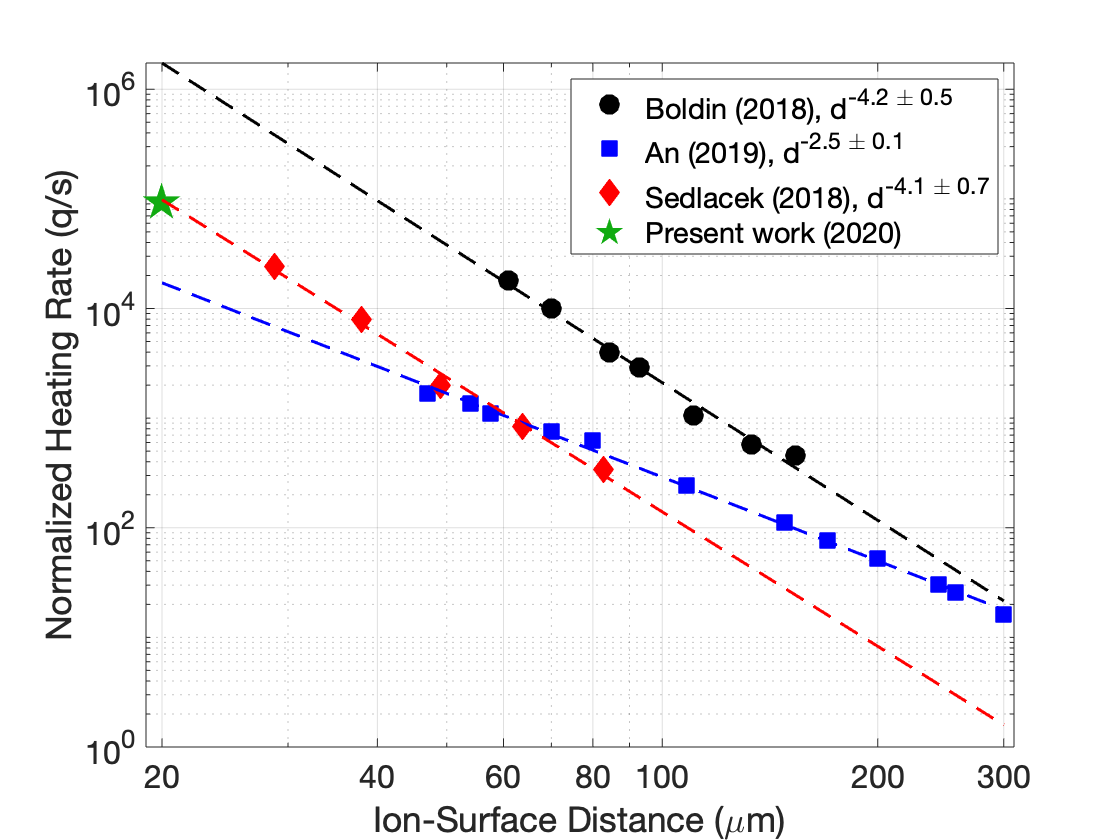}}
	\caption{Heating rate as a function of the ion-surface distance. The heating rate has been normalized for ion mass and secular motional trap frequency. We plot our data as the green star compared to previously published data from \cite{boldin:2018} (black circles), \cite{an:2019} (blue squares), and \cite{sedlacek:2018} (red diamonds). The scaling of each fit is indicated in the legend. Data reprinted with permission from authors of \cite{an:2019}.}
	\label{fig:HRvDist}
\end{figure}

We can normalize our measured heating rate for ion mass and trap frequency to compare with community data \cite{hite:2012, noel:2019, an:2019}. To do so, we consider the heating rate given by $\langle \dot{n}\rangle = \frac{q^2}{4m \hbar \omega}S_E$ where $q$ is the charge of the ion, $m$ is the mass of the ion, $\omega$ is the motional frequency, and $S_E$ is the spectral density of the electric field noise which goes like $1/\omega$. Scaling to $\omega =2\pi \times 1$~MHz and $^{40}$Ca$^+$ allows us to compare our data point at 20~$\mu$m ion-surface distance with published data at larger ion-surface distances. Fig \ref{fig:HRvDist} shows that our normalized heating rate falls along the extrapolated $d^{-4}$ scaling found in other experiments \cite{an:2019,sedlacek:2018, boldin:2018}.

\begin{figure}[htb]
	\resizebox{.45\textwidth}{!}{
	\includegraphics{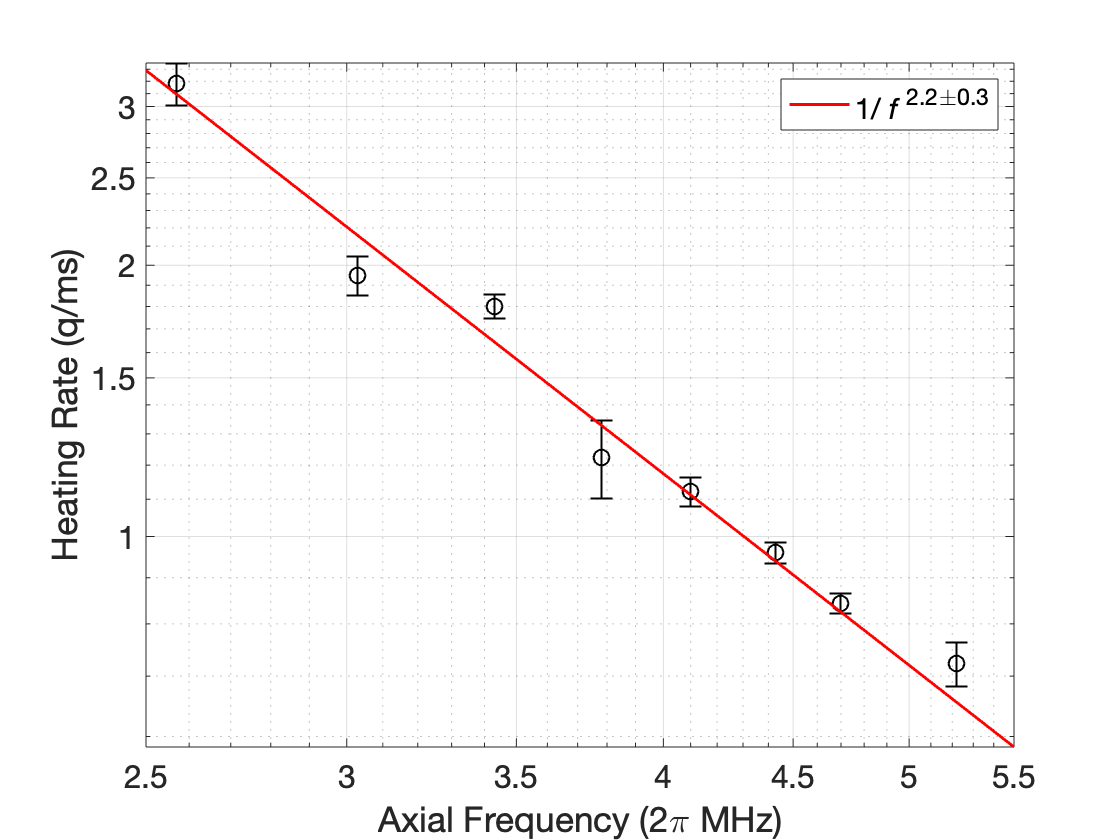}}
	\caption{Heating rate as a function of axial secular motional trap frequency follows a $1/f^{2.2\pm0.3}$ trend consistent with $1/f$ scaling of electric field spectral density \cite{sedlacek:2018, an:2019}.}
	\label{fig:heatingFreq}
\end{figure}

We also measure the heating rate as a function of axial secular trap frequency over the loading hole. Figure~\ref{fig:heatingFreq} shows our results as we scan the frequency from 2$\pi \times$2.5 to 2$\pi \times$5.2~MHz. The fitted line represents power law $\langle \dot{n} \rangle \propto 1/f^{2.2\pm0.3}$ which agrees well with previously reported heating rate trends and is consistent with $1/f$ scaling of electric field spectral density \cite{sedlacek:2018, an:2019}.

\subsection{Waveguide beam profile}
\label{WGprofile}

\begin{figure}[ht]
\resizebox{.45\textwidth}{!}{
\includegraphics{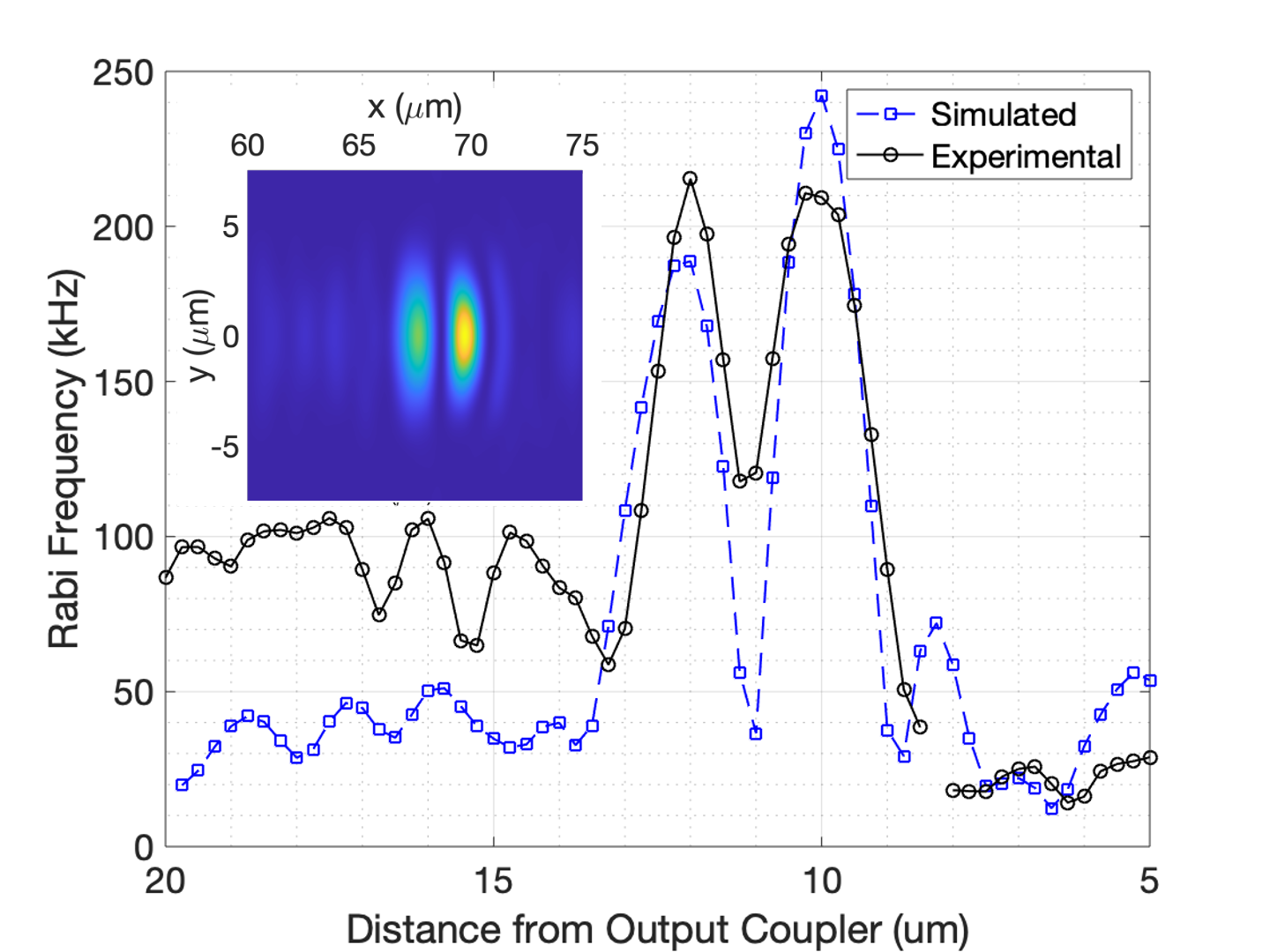}}
\caption{Measured (black) and simulated (blue) Rabi frequency of the $\Delta m_F = +1$ transition using the waveguide delivered beam as a function of the ion's distance from the loading hole. The two maxima, indicating highest beam intensities, occur when the ion is 10~$\mu$m and 12~$\mu$m from the output grating. The dip is attributed to variations in the distance between the grating and reflective layer and can be reproduced in simulations that vary this distance by 60 nm (blue). The simulated Rabi frequency is calculated from the simulated waveguide output intensity and is scaled to match the experimental Rabi frequency. (inset) The simulated 2D waveguide output intensity cross section at the ion height.}
\label{fig:intensityProfile}
\end{figure}

We couple 435~nm light into the waveguide via the input diffraction grating shown in Figure~\ref{fig:trapWG}d which emits light from the output grating at a measured angle of 66$^{\circ}$ (Figure~\ref{fig:trapWG}f). This differs from the simulated output angle of 63$^{\circ}$ due to fabrication tolerances.  Roughly 26 mW of power is incident on the input grating.
We determine the output intensity by measuring the Rabi rate of the ion at various positions around the expected grating output and plot the relative Rabi frequencies as a function of ion position for the $\Delta m_F = +1$ transition (see Figure~\ref{fig:intensityProfile}). In this device, we observed a double peak profile from the grating. Note the dip in transmission near 11~$\mu$m from the output grating. We attribute this to destructive interference in the output beam due to variations in the distance between the grating and reflector described in the Trap Design and Fabrication Section. Due to limitations in the fabrication tolerances, there are variations in the oxide thickness across the wafer, and therefore the actual distance between the grating and reflector varies across the wafer. This variation in spacing between grating and reflective surface is a result of the chemical mechanical polishing process which is used to polish the oxide layer above the metal, prior to deposition of the silicon nitride waveguide layer. This process results in variation of the oxide thickness across the wafer on the order of 100~nm. 

 This double-peak feature was reproduced in simulation by vertically shifting the location of the reflector by 60~nm from the nominal design point, which results in a reduction of the output efficiency from -1.6~dB in the ideal case to -3.7~dB, and two sub-micron peaks separated by 1.8~$\mu$m. The simulated output intensity profile with this 60~nm shift is shown as a 2D cross section at the anticipated ion height in the Figure~\ref{fig:intensityProfile} inset. The simulated 1D Rabi frequency corresponding to $y=0$~$\mu$m in the inset is shown in blue in Figure~\ref{fig:intensityProfile}.

In future devices the fabrication uncertainty will be reduced by altering the fabrication process such that the thickness of the oxide between the metal reflector and grating are set by a single deposition process, without a polishing step. This is expected to reduce cross-wafer oxide variation below 10~nm and eliminate the double peak in the output profile.

Despite this feature, we have observed $\pi$-times as fast as 4.13$\pm$0.02~$\mu$s corresponding to Rabi frequencies as high as $\Omega = 2\pi\times 121.1\pm0.6$~kHz when coupling to the waveguide is optimized. This corresponds to an estimated peak intensity of nearly 300~nW/$\mu$m$^2$ and through-chip coupling of -39 dB.

\subsection{Electrostatic charging measurements and results}

In some traps with integrated photonics, an ITO coating is deposited over the exposed dielectrics to prevent charge buildup \cite{niffenegger:2020}. For our traps, ITO was not employed as it was not compatible with the CMOS facility used at Sandia. Consequently, we observe an effect due to photo-liberated electrons charging the 35~$\mu$m$^2$ area of exposed dielectric material (corresponding to a 0.31 steradian solid angle) that composes the output grating coupler of the integrated waveguide. The electric field from the charged grating coupler simultaneously shifts the position of the ion and perturbs the harmonic potential of the prescribed trapping fields. We measure this effect by monitoring the axial motional sideband frequency in two different scenarios: 1) Charging is induced via continuously applied light through the waveguide (off-resonant) and the sideband frequency is probed via a free space beam; 2) The charging is induced \textit{and} the sideband frequency is probed via light coupled through the waveguide. In the latter case, the light through the waveguide is pulsed on only during the probe time, which is more representative of normal operation.

\subsubsection{Continuous Photo-induced Charging}

We first investigate the photo-induced charging effect at a long timescale by measuring the axial secular frequency with a free-space beam. Using a free space beam to probe the ion allows us to maintain a constant optical power through the waveguide over the duration of the measurement. The secular frequency is measured from the motional sideband of the $|2S_{1/2},F=0,m_F=0\rangle$ to $|2D_{3/2},F=2,m_F=+2\rangle$ transition. This measurement is performed with the ion directly above the grating (80~$\mu$m from the loading hole in Figure~\ref{fig:intensityProfile}). We begin with the waveguide-coupled light shuttered. We interrogate the resonant frequencies of the red and blue axial sidebands approximately every 15~seconds while the waveguide coupler is continuously charged for $2000$~s by turning on the non-resonant waveguide-coupled light and then discharged for $\> 2500$~s by turning off the waveguide-coupled light.

\begin{figure}[ht]
	\resizebox{.45\textwidth}{!}{
		\includegraphics{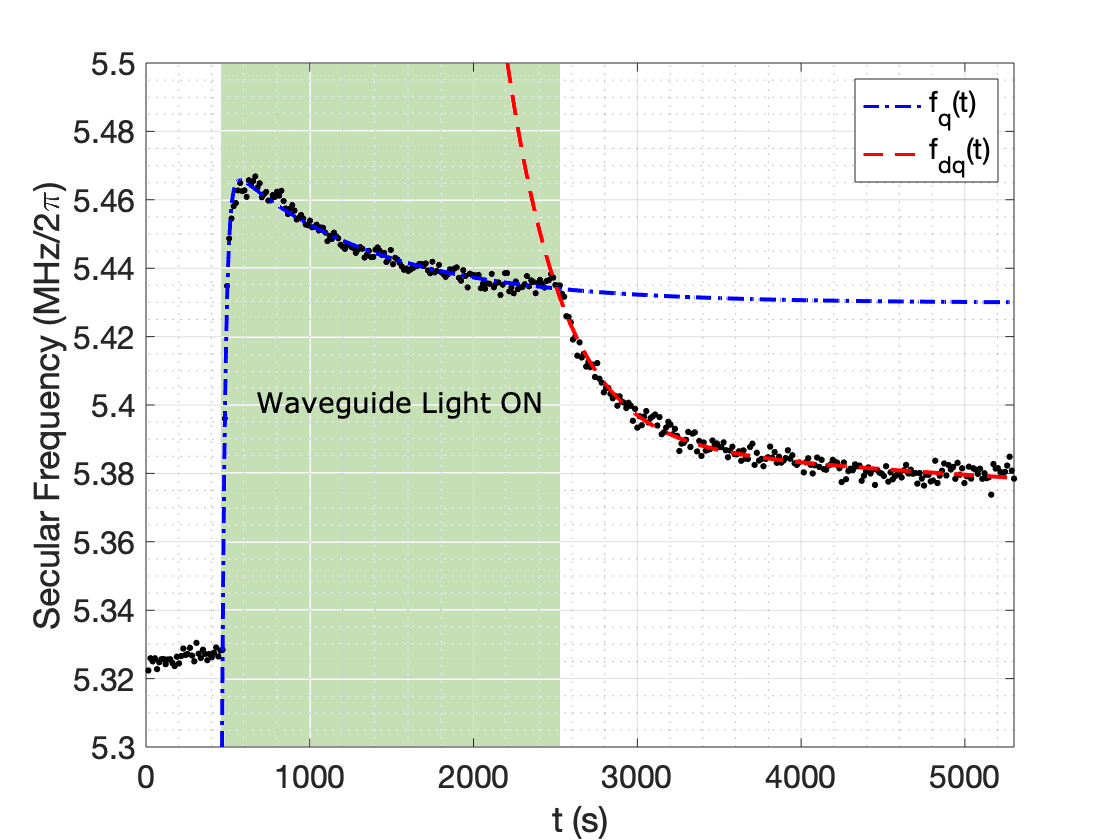}}
	\caption{Secular frequency vs time as waveguide-coupled light is turned on and off. Secular frequency is measured with a free-space beam tuned to the $\Delta m_F = +2$ transition. Between 500~s and 2500~s, the waveguide-coupled light is turned on (green) and servoed to a constant power. The two opposing charge effects are attributed to the metal and dielectric layers as described in \cite{harlander:2010}. Fitted lines for the charging (blue dot dash) and discharging (red dash) are described by Eqs. \ref{eq:q} and \ref{eq:dq}.}
	\label{fig:charging}
\end{figure}

The results of the charging experiment are shown in Figure~\ref{fig:charging}. The first $400$~s shows the baseline axial frequency after the waveguide has been off for a period of $\> 12$~hours. We turn on the waveguide-coupled light at ~400~s $f$. The charging is characterized by two main effects. The primary effect is a fast negative charging of the top metal layer. This is attributed to a small electric dipole created when electrons are excited from the metal onto insulating patches of oxide above \cite{harlander:2010}. This effect is superimposed by a slower secondary effect of positive charging of the dielectric grating. These two effects have been previously observed and are well characterized in similar photo-induced charge studies \cite{harlander:2010}. When the waveguide-coupled light is turned off at $2500$~s, both materials slowly discharge.

With the light on, the frequency shift due to charging $f_q(t)$ can be fit to the following form:
\begin{eqnarray}
f_q(t) = && \Delta f_1\left(1-\textrm{exp}\left(\frac{-(t-t_{on})}{T_1}\right)\right) \nonumber \\
&& - \Delta f_2\left(1-\textrm{exp}\left(\frac{-(t-t_{on})}{T_2}\right)\right) + f_0
\label{eq:q}
\end{eqnarray}
where $\Delta f_{1,2}$ are the settling frequency offsets of the two charging effects. The difference $\Delta f_1 - \Delta f_2$ represents the settling frequency offset with the waveguide light on. The terms $t_\textrm{on}$ and $T_{1,2}$ represent the turn on time of the waveguide light and the charging time, respectively. The initial frequency prior to turning on the waveguide light is $f_0$.

After the waveguide light has been extinguished, the frequency shift due to discharging $f_{dq}(t)$ can be described as follows:
\begin{eqnarray}
f_{dq}(t) = && \Delta f_3\left(-\textrm{exp}\left(\frac{-(t-t_{off})}{T_3}\right)\right) \nonumber \\
&& + \Delta f_4\left(-\textrm{exp}\left(\frac{-(t-t_{off})}{T_4}\right)\right) + f_0
\label{eq:dq}
\end{eqnarray}
where $\Delta f_{3,4}$ and $T_{3,4}$ are the settling frequency offsets and the discharging times of the two materials, respectively, and $t_\textrm{off}$ is the turn off time of the waveguide light. 

In Figure~\ref{fig:charging}, upon waveguide turn on, the charging happens over $T_1 = 21$~s and $T_2 \sim 900$~s and the difference of the settling frequency offsets is $\Delta f_1-\Delta f_2=101$~kHz. After the waveguide turns off, the discharging times are $T_3 \sim 6$~min and $T_4\sim 5$~hrs as the frequency is expected to return to its original $f_0$ value of $5.329$~MHz. We can negate the settled frequency offset $\Delta f_1-\Delta f_2\simeq 0.1$~MHz by applying a $\sim$2.4~kV/cm electric field in the vertical direction.

For this experiment, approximately 17~mW of off-resonant 435~nm light is directed on the input coupler to the waveguide. To better understand how much power is coupled through the waveguide, we direct the same amount of power of resonant 435~nm light onto the input coupler and we shuttle the ion 68~$\mu$m from the loading hole (see Figure~\ref{fig:intensityProfile}) where the waveguide outcoupled light is maximally overlapped with the ion. Similar to the beam profile experiments in the Waveguide beam profile subsection, we use the waveguide coupled light to perform Rabi flops on the $|2S_{1/2},F=0,m_F=0\rangle$ to $|2D_{3/2},F=2,m_F=+1\rangle$ transition and achieve $\pi$-times of 5~$\mu$s for this particular experiment. Due to the angle of the waveguide beam with respect to our magnetic field, the $\Delta m_F=+1$ transition is most strongly coupled by the waveguide beam.

\subsubsection{Charging Due to Pulsed Operation}

\begin{figure}[ht]
	\resizebox{.45\textwidth}{!}{
		\includegraphics{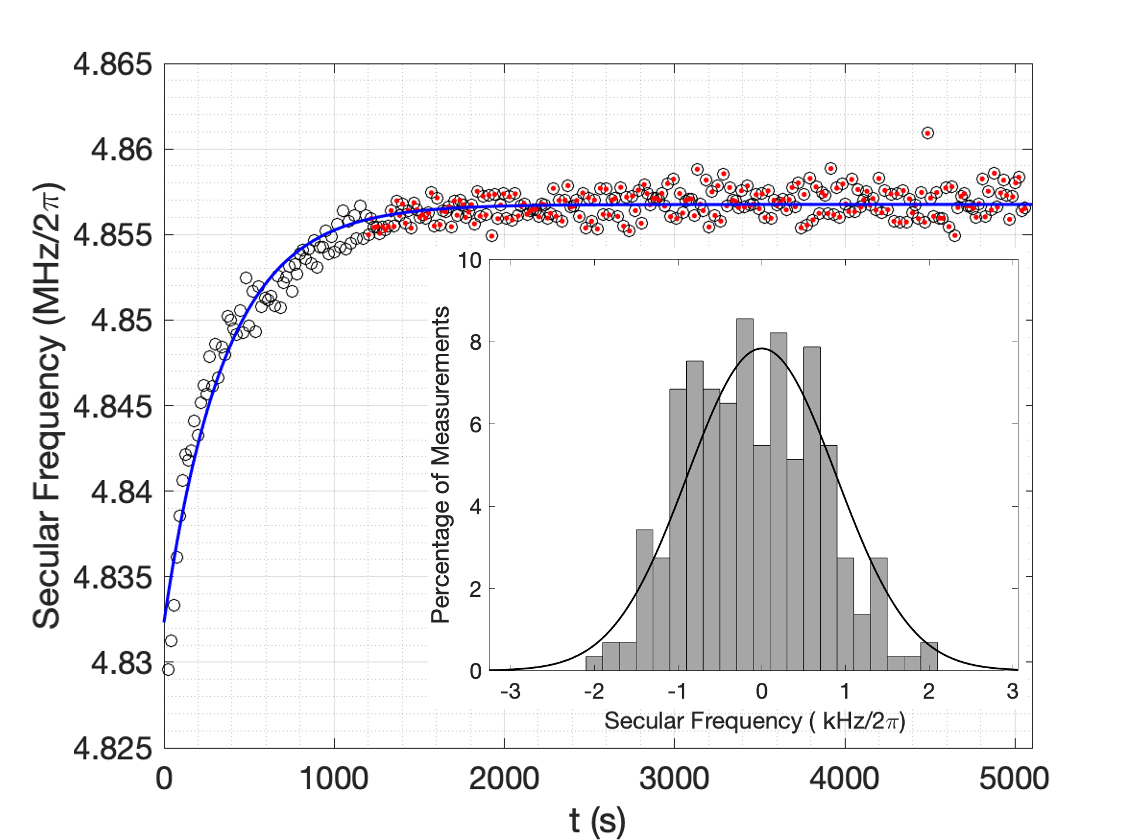}}
	\caption{Secular frequency over time during typical clock operation, as measured by the waveguide coupled light tuned to the $\Delta m_F=1$ transition. (inset) Histogram of the settled frequency (red points) difference from the fitted curve (blue line). The histogram is fitted to a Gaussian with $\sigma =0.9$~kHz.}
	\label{fig:chargestability}
\end{figure}

We also monitor the charge-induced frequency shifts using the waveguide coupled light rather than the free-space light in order to determine the stability of the shifted frequency during more typical clock operation. Figure~\ref{fig:chargestability} shows the frequency shift of the $\Delta m_F=1$ transition during a typical duty cycle. The charging happens on a much slower timescale because the light is only on for a fraction of the time as the measurements in Figure~\ref{fig:charging}.  In these experiments, the waveguide coupled light is only on during probe time for 15 $\mu$s, which contributes $0.61\%$ of the duty cycle. A histogram of the difference between the fitted exponential and the data points is shown in the inset for the red points, with a standard deviation of $\sigma =0.9$~kHz. 

\section{Conclusion}
\label{sec:conclusion}

We have fabricated a surface trap with integrated waveguides for delivering 435~nm light to address the quadrupole transition in \ytterbium. To assess the feasibility of dielectric waveguides in quantum applications, we have characterized the heating rate as a function of proximity to exposed dielectric output grating and of axial secular trap frequency. We see no increase in heating rate over the dielectric, despite having an ion-surface distance of only 20~$\mu$m. The heating rate scaling with axial secular trap frequency follows a $1/f^{2.2}$ trend consistent with previous results. 

We also mapped the output profile of the waveguide by measuring the Rabi frequency of the ion with the waveguide delivered light and showed that it is consistent with simulations and trap fabrication tolerances. We anticipate the beam profile to significantly improve in future devices by eliminating the double-peak feature present in the current device output. Furthermore, transitioning to an edge coupled instead of diffraction grating coupled device at the input is anticipated to improve the input coupling to -3~dB or better, depending on the process used. Despite this feature, we have measured $\pi$-times as fast as 4.13$\pm$0.02~$\mu$s. 

Finally, we have characterized the electric field shift and stability due to the photo-induced charging of the trap. The lack of deleterious heating over exposed dielectrics or destabilizing photo-induced electric fields is a promising result for the future utilization of these devices in quantum computing and timekeeping applications.

\begin{acknowledgments}

The authors thank the members of Sandia's Microsystems and Engineering Sciences Application (MESA) facility for their fabrication expertise for helpful comments on the manuscript, as well as H. Haeffner and D. An for their discussions and data for Figure~\ref{fig:HRvDist}. This work was supported by the Defense Advanced Research Projects Activity (DARPA).
Sandia National Laboratories is a multimission laboratory managed and operated by National Technology \& Engineering Solutions of Sandia, LLC, a wholly owned subsidiary of Honeywell International Inc., for the U.S. Department of Energy's National Nuclear Security Administration under contract DE-NA0003525. This paper describes objective technical results and analysis. Any subjective views or opinions that might be expressed in the paper do not necessarily represent the views of the U.S. Department of Energy or the United States Government. 

\end{acknowledgments}


\begin{thebibliography}{25}%
	\makeatletter
	\providecommand \@ifxundefined [1]{%
		\@ifx{#1\undefined}
	}%
	\providecommand \@ifnum [1]{%
		\ifnum #1\expandafter \@firstoftwo
		\else \expandafter \@secondoftwo
		\fi
	}%
	\providecommand \@ifx [1]{%
		\ifx #1\expandafter \@firstoftwo
		\else \expandafter \@secondoftwo
		\fi
	}%
	\providecommand \natexlab [1]{#1}%
	\providecommand \enquote  [1]{``#1''}%
	\providecommand \bibnamefont  [1]{#1}%
	\providecommand \bibfnamefont [1]{#1}%
	\providecommand \citenamefont [1]{#1}%
	\providecommand \href@noop [0]{\@secondoftwo}%
	\providecommand \href [0]{\begingroup \@sanitize@url \@href}%
	\providecommand \@href[1]{\@@startlink{#1}\@@href}%
	\providecommand \@@href[1]{\endgroup#1\@@endlink}%
	\providecommand \@sanitize@url [0]{\catcode `\\12\catcode `\$12\catcode
		`\&12\catcode `\#12\catcode `\^12\catcode `\_12\catcode `\%12\relax}%
	\providecommand \@@startlink[1]{}%
	\providecommand \@@endlink[0]{}%
	\providecommand \url  [0]{\begingroup\@sanitize@url \@url }%
	\providecommand \@url [1]{\endgroup\@href {#1}{\urlprefix }}%
	\providecommand \urlprefix  [0]{URL }%
	\providecommand \Eprint [0]{\href }%
	\providecommand \doibase [0]{https://doi.org/}%
	\providecommand \selectlanguage [0]{\@gobble}%
	\providecommand \bibinfo  [0]{\@secondoftwo}%
	\providecommand \bibfield  [0]{\@secondoftwo}%
	\providecommand \translation [1]{[#1]}%
	\providecommand \BibitemOpen [0]{}%
	\providecommand \bibitemStop [0]{}%
	\providecommand \bibitemNoStop [0]{.\EOS\space}%
	\providecommand \EOS [0]{\spacefactor3000\relax}%
	\providecommand \BibitemShut  [1]{\csname bibitem#1\endcsname}%
	\let\auto@bib@innerbib\@empty
	\bibitem [{\citenamefont {Stick}\ \emph {et~al.}(2005)\citenamefont {Stick},
		\citenamefont {Hensinger}, \citenamefont {Olmschenk}, \citenamefont {Madsen},
		\citenamefont {Schwab},\ and\ \citenamefont {Monroe}}]{stick:2005}%
	\BibitemOpen
	\bibfield  {author} {\bibinfo {author} {\bibfnamefont {D.}~\bibnamefont
			{Stick}}, \bibinfo {author} {\bibfnamefont {W.~K.}\ \bibnamefont
			{Hensinger}}, \bibinfo {author} {\bibfnamefont {S.}~\bibnamefont
			{Olmschenk}}, \bibinfo {author} {\bibfnamefont {M.}~\bibnamefont {Madsen}},
		\bibinfo {author} {\bibfnamefont {K.}~\bibnamefont {Schwab}},\ and\ \bibinfo
		{author} {\bibfnamefont {C.}~\bibnamefont {Monroe}},\ }\bibfield  {title}
	{\bibinfo {title} {Ion trap in a semiconductor chip},\ }\href
	{https://doi.org/10.1038/nphys171} {\bibfield  {journal} {\bibinfo  {journal}
			{Nat. Phys.}\ }\textbf {\bibinfo {volume} {2}},\ \bibinfo {pages} {36}
		(\bibinfo {year} {2005})}\BibitemShut {NoStop}%
	\bibitem [{\citenamefont {Seidelin}\ \emph {et~al.}(2006)\citenamefont
		{Seidelin}, \citenamefont {Chiaverini}, \citenamefont {Reichle},
		\citenamefont {Bollinger}, \citenamefont {Leibfried}, \citenamefont
		{Britton}, \citenamefont {Wesenberg}, \citenamefont {Blakestad},
		\citenamefont {Epstein}, \citenamefont {Hume}, \citenamefont {Itano},
		\citenamefont {Jost}, \citenamefont {Langer}, \citenamefont {Ozeri},
		\citenamefont {Shiga},\ and\ \citenamefont {Wineland}}]{seidelin:2006}%
	\BibitemOpen
	\bibfield  {author} {\bibinfo {author} {\bibfnamefont {S.}~\bibnamefont
			{Seidelin}}, \bibinfo {author} {\bibfnamefont {J.}~\bibnamefont
			{Chiaverini}}, \bibinfo {author} {\bibfnamefont {R.}~\bibnamefont {Reichle}},
		\bibinfo {author} {\bibfnamefont {J.~J.}\ \bibnamefont {Bollinger}}, \bibinfo
		{author} {\bibfnamefont {D.}~\bibnamefont {Leibfried}}, \bibinfo {author}
		{\bibfnamefont {J.}~\bibnamefont {Britton}}, \bibinfo {author} {\bibfnamefont
			{J.~H.}\ \bibnamefont {Wesenberg}}, \bibinfo {author} {\bibfnamefont {R.~B.}\
			\bibnamefont {Blakestad}}, \bibinfo {author} {\bibfnamefont {R.~J.}\
			\bibnamefont {Epstein}}, \bibinfo {author} {\bibfnamefont {D.~B.}\
			\bibnamefont {Hume}}, \bibinfo {author} {\bibfnamefont {W.~M.}\ \bibnamefont
			{Itano}}, \bibinfo {author} {\bibfnamefont {J.~D.}\ \bibnamefont {Jost}},
		\bibinfo {author} {\bibfnamefont {C.}~\bibnamefont {Langer}}, \bibinfo
		{author} {\bibfnamefont {R.}~\bibnamefont {Ozeri}}, \bibinfo {author}
		{\bibfnamefont {N.}~\bibnamefont {Shiga}},\ and\ \bibinfo {author}
		{\bibfnamefont {D.~J.}\ \bibnamefont {Wineland}},\ }\bibfield  {title}
	{\bibinfo {title} {Microfabricated surface-electrode ion trap for scalable
			quantum information processing},\ }\href
	{https://doi.org/10.1103/PhysRevLett.96.253003} {\bibfield  {journal}
		{\bibinfo  {journal} {Phys. Rev. Lett.}\ }\textbf {\bibinfo {volume} {96}},\
		\bibinfo {pages} {253003} (\bibinfo {year} {2006})}\BibitemShut {NoStop}%
	\bibitem [{\citenamefont {Brewer}\ \emph
		{et~al.}(2019{\natexlab{a}})\citenamefont {Brewer}, \citenamefont {Chen},
		\citenamefont {Hankin}, \citenamefont {Clements}, \citenamefont {Chou},
		\citenamefont {Wineland}, \citenamefont {Hume},\ and\ \citenamefont
		{Leibrandt}}]{brewer:2019a}%
	\BibitemOpen
	\bibfield  {author} {\bibinfo {author} {\bibfnamefont {S.~M.}\ \bibnamefont
			{Brewer}}, \bibinfo {author} {\bibfnamefont {J.-S.}\ \bibnamefont {Chen}},
		\bibinfo {author} {\bibfnamefont {A.~M.}\ \bibnamefont {Hankin}}, \bibinfo
		{author} {\bibfnamefont {E.~R.}\ \bibnamefont {Clements}}, \bibinfo {author}
		{\bibfnamefont {C.~W.}\ \bibnamefont {Chou}}, \bibinfo {author}
		{\bibfnamefont {D.~J.}\ \bibnamefont {Wineland}}, \bibinfo {author}
		{\bibfnamefont {D.~B.}\ \bibnamefont {Hume}},\ and\ \bibinfo {author}
		{\bibfnamefont {D.~R.}\ \bibnamefont {Leibrandt}},\ }\bibfield  {title}
	{\bibinfo {title} {$^{27}{\mathrm{al}}^{+}$ quantum-logic clock with a
			systematic uncertainty below ${10}^{\ensuremath{-}18}$},\ }\href
	{https://doi.org/10.1103/PhysRevLett.123.033201} {\bibfield  {journal}
		{\bibinfo  {journal} {Phys. Rev. Lett.}\ }\textbf {\bibinfo {volume} {123}},\
		\bibinfo {pages} {033201} (\bibinfo {year} {2019}{\natexlab{a}})}\BibitemShut
	{NoStop}%
	\bibitem [{\citenamefont {Brewer}\ \emph
		{et~al.}(2019{\natexlab{b}})\citenamefont {Brewer}, \citenamefont {Chen},
		\citenamefont {Beloy}, \citenamefont {Hankin}, \citenamefont {Clements},
		\citenamefont {Chou}, \citenamefont {McGrew}, \citenamefont {Zhang},
		\citenamefont {Fasano}, \citenamefont {Nicolodi}, \citenamefont {Leopardi},
		\citenamefont {Fortier}, \citenamefont {Diddams}, \citenamefont {Ludlow},
		\citenamefont {Wineland}, \citenamefont {Leibrandt},\ and\ \citenamefont
		{Hume}}]{brewer:2019b}%
	\BibitemOpen
	\bibfield  {author} {\bibinfo {author} {\bibfnamefont {S.~M.}\ \bibnamefont
			{Brewer}}, \bibinfo {author} {\bibfnamefont {J.-S.}\ \bibnamefont {Chen}},
		\bibinfo {author} {\bibfnamefont {K.}~\bibnamefont {Beloy}}, \bibinfo
		{author} {\bibfnamefont {A.~M.}\ \bibnamefont {Hankin}}, \bibinfo {author}
		{\bibfnamefont {E.~R.}\ \bibnamefont {Clements}}, \bibinfo {author}
		{\bibfnamefont {C.~W.}\ \bibnamefont {Chou}}, \bibinfo {author}
		{\bibfnamefont {W.~F.}\ \bibnamefont {McGrew}}, \bibinfo {author}
		{\bibfnamefont {X.}~\bibnamefont {Zhang}}, \bibinfo {author} {\bibfnamefont
			{R.~J.}\ \bibnamefont {Fasano}}, \bibinfo {author} {\bibfnamefont
			{D.}~\bibnamefont {Nicolodi}}, \bibinfo {author} {\bibfnamefont
			{H.}~\bibnamefont {Leopardi}}, \bibinfo {author} {\bibfnamefont {T.~M.}\
			\bibnamefont {Fortier}}, \bibinfo {author} {\bibfnamefont {S.~A.}\
			\bibnamefont {Diddams}}, \bibinfo {author} {\bibfnamefont {A.~D.}\
			\bibnamefont {Ludlow}}, \bibinfo {author} {\bibfnamefont {D.~J.}\
			\bibnamefont {Wineland}}, \bibinfo {author} {\bibfnamefont {D.~R.}\
			\bibnamefont {Leibrandt}},\ and\ \bibinfo {author} {\bibfnamefont {D.~B.}\
			\bibnamefont {Hume}},\ }\bibfield  {title} {\bibinfo {title} {Measurements of
			$^{27}\mathrm{Al}^{+}$ and $^{25}\mathrm{Mg}^{+}$ magnetic constants for
			improved ion-clock accuracy},\ }\href
	{https://doi.org/10.1103/PhysRevA.100.013409} {\bibfield  {journal} {\bibinfo
			{journal} {Phys. Rev. A}\ }\textbf {\bibinfo {volume} {100}},\ \bibinfo
		{pages} {013409} (\bibinfo {year} {2019}{\natexlab{b}})}\BibitemShut
	{NoStop}%
	\bibitem [{\citenamefont {Borregaard}\ and\ \citenamefont
		{Sorensen}(2013)}]{borregaard:2013}%
	\BibitemOpen
	\bibfield  {author} {\bibinfo {author} {\bibfnamefont {J.}~\bibnamefont
			{Borregaard}}\ and\ \bibinfo {author} {\bibfnamefont {A.S.}~\bibnamefont
			{Sorensen}},\ }\bibfield  {title} {\bibinfo {title} {Efficient atomic clocks
			operated with several atomic ensembles},\ }\href@noop {} {\bibfield
		{journal} {\bibinfo  {journal} {Phys. Rev. Lett.}\ }\textbf {\bibinfo
			{volume} {111}},\ \bibinfo {pages} {090802} (\bibinfo {year}
		{2013})}\BibitemShut {NoStop}%
	\bibitem [{\citenamefont {Mehta}\ \emph {et~al.}(2016)\citenamefont {Mehta},
		\citenamefont {Bruzewicz}, \citenamefont {McConnell}, \citenamefont {Ram},
		\citenamefont {Sage},\ and\ \citenamefont {Chiaverini}}]{mehta:2016}%
	\BibitemOpen
	\bibfield  {author} {\bibinfo {author} {\bibfnamefont {K.}~\bibnamefont
			{Mehta}}, \bibinfo {author} {\bibfnamefont {C.}~\bibnamefont {Bruzewicz}},
		\bibinfo {author} {\bibfnamefont {R.}~\bibnamefont {McConnell}}, \bibinfo
		{author} {\bibfnamefont {R.~J.}\ \bibnamefont {Ram}}, \bibinfo {author}
		{\bibfnamefont {J.~M.}\ \bibnamefont {Sage}},\ and\ \bibinfo {author}
		{\bibfnamefont {J.}~\bibnamefont {Chiaverini}},\ }\bibfield  {title}
	{\bibinfo {title} {Integrated optical addressing of an ion qubit},\
	}\href@noop {} {\bibfield  {journal} {\bibinfo  {journal} {Nature
				Nanotechnology}\ }\textbf {\bibinfo {volume} {11}},\ \bibinfo {pages} {1066 }
		(\bibinfo {year} {2016})}\BibitemShut {NoStop}%
	\bibitem [{\citenamefont {Jiang}\ \emph {et~al.}(2011)\citenamefont {Jiang},
		\citenamefont {Whitten},\ and\ \citenamefont {Pau}}]{jiang:2011}%
	\BibitemOpen
	\bibfield  {author} {\bibinfo {author} {\bibfnamefont {L.}~\bibnamefont
			{Jiang}}, \bibinfo {author} {\bibfnamefont {W.~B.}\ \bibnamefont {Whitten}},\
		and\ \bibinfo {author} {\bibfnamefont {S.}~\bibnamefont {Pau}},\ }\bibfield
	{title} {\bibinfo {title} {A planar ion trapping microdevice with integrated
			waveguides for optical detection},\ }\href
	{https://doi.org/10.1364/OE.19.003037} {\bibfield  {journal} {\bibinfo
			{journal} {Opt. Express}\ }\textbf {\bibinfo {volume} {19}},\ \bibinfo
		{pages} {3037} (\bibinfo {year} {2011})}\BibitemShut {NoStop}%
	\bibitem [{\citenamefont {Slichter}\ \emph {et~al.}(2017)\citenamefont
		{Slichter}, \citenamefont {Verma}, \citenamefont {Leibfried}, \citenamefont
		{Mirin}, \citenamefont {Nam},\ and\ \citenamefont
		{Wineland}}]{slichter:2017}%
	\BibitemOpen
	\bibfield  {author} {\bibinfo {author} {\bibfnamefont {D.~H.}\ \bibnamefont
			{Slichter}}, \bibinfo {author} {\bibfnamefont {V.~B.}\ \bibnamefont {Verma}},
		\bibinfo {author} {\bibfnamefont {D.}~\bibnamefont {Leibfried}}, \bibinfo
		{author} {\bibfnamefont {R.~P.}\ \bibnamefont {Mirin}}, \bibinfo {author}
		{\bibfnamefont {S.~W.}\ \bibnamefont {Nam}},\ and\ \bibinfo {author}
		{\bibfnamefont {D.~J.}\ \bibnamefont {Wineland}},\ }\bibfield  {title}
	{\bibinfo {title} {Uv-sensitive superconducting nanowire single photon
			detectors for integration in an ion trap},\ }\href
	{https://doi.org/10.1364/OE.25.008705} {\bibfield  {journal} {\bibinfo
			{journal} {Opt. Express}\ }\textbf {\bibinfo {volume} {25}},\ \bibinfo
		{pages} {8705} (\bibinfo {year} {2017})}\BibitemShut {NoStop}%
	\bibitem [{\citenamefont {Mehta}\ \emph {et~al.}(2020)\citenamefont {Mehta},
		\citenamefont {Zhang}, \citenamefont {Malinowski}, \citenamefont {Nguyen},
		\citenamefont {Stadler},\ and\ \citenamefont {Home}}]{mehta:2020}%
	\BibitemOpen
	\bibfield  {author} {\bibinfo {author} {\bibfnamefont {K.}~\bibnamefont
			{Mehta}}, \bibinfo {author} {\bibfnamefont {C.}~\bibnamefont {Zhang}},
		\bibinfo {author} {\bibfnamefont {M.}~\bibnamefont {Malinowski}}, \bibinfo
		{author} {\bibfnamefont {T.-L.}\ \bibnamefont {Nguyen}}, \bibinfo {author}
		{\bibfnamefont {M.}~\bibnamefont {Stadler}},\ and\ \bibinfo {author}
		{\bibfnamefont {J.~P.}\ \bibnamefont {Home}},\ }\bibfield  {title} {\bibinfo
		{title} {Integrated optical multi-ion quantum logic},\ }\href@noop {}
	{\bibfield  {journal} {\bibinfo  {journal} {Nature}\ }\textbf {\bibinfo
			{volume} {586}},\ \bibinfo {pages} {533} (\bibinfo {year}
		{2020})}\BibitemShut {NoStop}%
	\bibitem [{\citenamefont {West}\ \emph {et~al.}(1998)\citenamefont {West},
		\citenamefont {Loh}, \citenamefont {Kharas}, \citenamefont {SoraceAgaskar},
		\citenamefont {Mehta}, \citenamefont {Sage}, \citenamefont {Chiaverini},\
		and\ \citenamefont {Ram}}]{west:2019}%
	\BibitemOpen
	\bibfield  {author} {\bibinfo {author} {\bibfnamefont {G.~N.}\ \bibnamefont
			{West}}, \bibinfo {author} {\bibfnamefont {W.}~\bibnamefont {Loh}}, \bibinfo
		{author} {\bibfnamefont {D.}~\bibnamefont {Kharas}}, \bibinfo {author}
		{\bibfnamefont {C.}~\bibnamefont {SoraceAgaskar}}, \bibinfo {author}
		{\bibfnamefont {K.~K.}\ \bibnamefont {Mehta}}, \bibinfo {author}
		{\bibfnamefont {J.}~\bibnamefont {Sage}}, \bibinfo {author} {\bibfnamefont
			{J.}~\bibnamefont {Chiaverini}},\ and\ \bibinfo {author} {\bibfnamefont
			{R.~J.}\ \bibnamefont {Ram}},\ }\bibfield  {title} {\bibinfo {title}
		{Low-loss integrated photonics for the blue and ultraviolet regime},\
	}\href@noop {} {\bibfield  {journal} {\bibinfo  {journal} {APL Photonics}\
		}\textbf {\bibinfo {volume} {4}},\ \bibinfo {pages} {026101} (\bibinfo {year}
		{1998})}\BibitemShut {NoStop}%
	\bibitem [{\citenamefont {Sorace-Agaskar}\ \emph {et~al.}(2019)\citenamefont
		{Sorace-Agaskar}, \citenamefont {Kharas}, \citenamefont {Yegnanarayanan},
		\citenamefont {Maxson}, \citenamefont {West}, \citenamefont {Loh},
		\citenamefont {Bramhavar}, \citenamefont {Ram}, \citenamefont {Chiaverini},
		\citenamefont {Sage},\ and\ \citenamefont {Juodawlkis}}]{sorace:2019}%
	\BibitemOpen
	\bibfield  {author} {\bibinfo {author} {\bibfnamefont {C.}~\bibnamefont
			{Sorace-Agaskar}}, \bibinfo {author} {\bibfnamefont {D.}~\bibnamefont
			{Kharas}}, \bibinfo {author} {\bibfnamefont {S.}~\bibnamefont
			{Yegnanarayanan}}, \bibinfo {author} {\bibfnamefont {R.~T.}\ \bibnamefont
			{Maxson}}, \bibinfo {author} {\bibfnamefont {G.~N.}\ \bibnamefont {West}},
		\bibinfo {author} {\bibfnamefont {W.}~\bibnamefont {Loh}}, \bibinfo {author}
		{\bibfnamefont {S.}~\bibnamefont {Bramhavar}}, \bibinfo {author}
		{\bibfnamefont {R.~J.}\ \bibnamefont {Ram}}, \bibinfo {author} {\bibfnamefont
			{J.}~\bibnamefont {Chiaverini}}, \bibinfo {author} {\bibfnamefont
			{J.}~\bibnamefont {Sage}},\ and\ \bibinfo {author} {\bibfnamefont
			{P.}~\bibnamefont {Juodawlkis}},\ }\bibfield  {title} {\bibinfo {title}
		{Versatile silicon nitride and alumina integratedphotonic platforms for the
			ultravioletto short-wave infrared},\ }\href
	{https://doi.org/10.1109/JSTQE.2019.2904443} {\bibfield  {journal} {\bibinfo
			{journal} {IEEE Journal of Selected Topics in Quantum Electronics}\ }\textbf
		{\bibinfo {volume} {25}},\ \bibinfo {pages} {8201515} (\bibinfo {year}
		{2019})}\BibitemShut {NoStop}%
	\bibitem [{\citenamefont {Niffenegger}\ \emph {et~al.}(2020)\citenamefont
		{Niffenegger}, \citenamefont {Stuart}, \citenamefont {Sorace-Agaskar},
		\citenamefont {Kharas}, \citenamefont {Bramhavar}, \citenamefont {Bruzewicz},
		\citenamefont {Loh}, \citenamefont {McConnell}, \citenamefont {Reens},
		\citenamefont {West}, \citenamefont {Sage},\ and\ \citenamefont
		{Chiaverini}}]{niffenegger:2020}%
	\BibitemOpen
	\bibfield  {author} {\bibinfo {author} {\bibfnamefont {R.~J.}\ \bibnamefont
			{Niffenegger}}, \bibinfo {author} {\bibfnamefont {J.}~\bibnamefont {Stuart}},
		\bibinfo {author} {\bibfnamefont {C.}~\bibnamefont {Sorace-Agaskar}},
		\bibinfo {author} {\bibfnamefont {D.}~\bibnamefont {Kharas}}, \bibinfo
		{author} {\bibfnamefont {S.}~\bibnamefont {Bramhavar}}, \bibinfo {author}
		{\bibfnamefont {C.~D.}\ \bibnamefont {Bruzewicz}}, \bibinfo {author}
		{\bibfnamefont {W.}~\bibnamefont {Loh}}, \bibinfo {author} {\bibfnamefont
			{R.}~\bibnamefont {McConnell}}, \bibinfo {author} {\bibfnamefont
			{D.}~\bibnamefont {Reens}}, \bibinfo {author} {\bibfnamefont {G.~N.}\
			\bibnamefont {West}}, \bibinfo {author} {\bibfnamefont {J.~M.}\ \bibnamefont
			{Sage}},\ and\ \bibinfo {author} {\bibfnamefont {J.}~\bibnamefont
			{Chiaverini}},\ }\bibfield  {title} {\bibinfo {title} {Integrated
			multi-wavelength control of an ion qubit},\ }\href@noop {} {\bibfield
		{journal} {\bibinfo  {journal} {Nature}\ }\textbf {\bibinfo {volume} {586}},\
		\bibinfo {pages} {538} (\bibinfo {year} {2020})}\BibitemShut {NoStop}%
	\bibitem [{\citenamefont {Brownnutt}\ \emph {et~al.}(2015)\citenamefont
		{Brownnutt}, \citenamefont {Kumph}, \citenamefont {Rabl},\ and\ \citenamefont
		{Blatt}}]{brownnutt:2015}%
	\BibitemOpen
	\bibfield  {author} {\bibinfo {author} {\bibfnamefont {M.}~\bibnamefont
			{Brownnutt}}, \bibinfo {author} {\bibfnamefont {M.}~\bibnamefont {Kumph}},
		\bibinfo {author} {\bibfnamefont {P.}~\bibnamefont {Rabl}},\ and\ \bibinfo
		{author} {\bibfnamefont {R.}~\bibnamefont {Blatt}},\ }\bibfield  {title}
	{\bibinfo {title} {Ion-trap measurements of electric-field noise near
			surfaces},\ }\href {https://doi.org/10.1103/RevModPhys.87.1419} {\bibfield
		{journal} {\bibinfo  {journal} {Rev. Mod. Phys.}\ }\textbf {\bibinfo {volume}
			{87}},\ \bibinfo {pages} {1419} (\bibinfo {year} {2015})}\BibitemShut
	{NoStop}%
	\bibitem [{\citenamefont {Boldin}\ \emph {et~al.}(2018)\citenamefont {Boldin},
		\citenamefont {Kraft},\ and\ \citenamefont {Wunderlich}}]{boldin:2018}%
	\BibitemOpen
	\bibfield  {author} {\bibinfo {author} {\bibfnamefont {I.~A.}\ \bibnamefont
			{Boldin}}, \bibinfo {author} {\bibfnamefont {A.}~\bibnamefont {Kraft}},\ and\
		\bibinfo {author} {\bibfnamefont {C.}~\bibnamefont {Wunderlich}},\ }\bibfield
	{title} {\bibinfo {title} {Measuring anomalous heating in a planar ion trap
			with variable ion-surface separation},\ }\href
	{https://doi.org/10.1103/PhysRevLett.120.023201} {\bibfield  {journal}
		{\bibinfo  {journal} {Phys. Rev. Lett.}\ }\textbf {\bibinfo {volume} {120}},\
		\bibinfo {pages} {023201} (\bibinfo {year} {2018})}\BibitemShut {NoStop}%
	\bibitem [{\citenamefont {Sedlacek}\ \emph {et~al.}(2018)\citenamefont
		{Sedlacek}, \citenamefont {Greene}, \citenamefont {Stuart}, \citenamefont
		{McConnell}, \citenamefont {Bruzewicz}, \citenamefont {Sage},\ and\
		\citenamefont {Chiaverini}}]{sedlacek:2018}%
	\BibitemOpen
	\bibfield  {author} {\bibinfo {author} {\bibfnamefont {J.~A.}\ \bibnamefont
			{Sedlacek}}, \bibinfo {author} {\bibfnamefont {A.}~\bibnamefont {Greene}},
		\bibinfo {author} {\bibfnamefont {J.}~\bibnamefont {Stuart}}, \bibinfo
		{author} {\bibfnamefont {R.}~\bibnamefont {McConnell}}, \bibinfo {author}
		{\bibfnamefont {C.~D.}\ \bibnamefont {Bruzewicz}}, \bibinfo {author}
		{\bibfnamefont {J.~M.}\ \bibnamefont {Sage}},\ and\ \bibinfo {author}
		{\bibfnamefont {J.}~\bibnamefont {Chiaverini}},\ }\bibfield  {title}
	{\bibinfo {title} {Distance scaling of electric-field noise in a
			surface-electrode ion trap},\ }\href
	{https://doi.org/10.1103/PhysRevA.97.020302} {\bibfield  {journal} {\bibinfo
			{journal} {Phys. Rev. A}\ }\textbf {\bibinfo {volume} {97}},\ \bibinfo
		{pages} {020302(R)} (\bibinfo {year} {2018})}\BibitemShut {NoStop}%
	\bibitem [{\citenamefont {Hite}\ \emph {et~al.}(2012)\citenamefont {Hite},
		\citenamefont {Colombe}, \citenamefont {Wilson}, \citenamefont {Brown},
		\citenamefont {Warring}, \citenamefont {J\"ordens}, \citenamefont {Jost},
		\citenamefont {McKay}, \citenamefont {Pappas}, \citenamefont {Leibfried},\
		and\ \citenamefont {Wineland}}]{hite:2012}%
	\BibitemOpen
	\bibfield  {author} {\bibinfo {author} {\bibfnamefont {D.~A.}\ \bibnamefont
			{Hite}}, \bibinfo {author} {\bibfnamefont {Y.}~\bibnamefont {Colombe}},
		\bibinfo {author} {\bibfnamefont {A.~C.}\ \bibnamefont {Wilson}}, \bibinfo
		{author} {\bibfnamefont {K.~R.}\ \bibnamefont {Brown}}, \bibinfo {author}
		{\bibfnamefont {U.}~\bibnamefont {Warring}}, \bibinfo {author} {\bibfnamefont
			{R.}~\bibnamefont {J\"ordens}}, \bibinfo {author} {\bibfnamefont {J.~D.}\
			\bibnamefont {Jost}}, \bibinfo {author} {\bibfnamefont {K.~S.}\ \bibnamefont
			{McKay}}, \bibinfo {author} {\bibfnamefont {D.~P.}\ \bibnamefont {Pappas}},
		\bibinfo {author} {\bibfnamefont {D.}~\bibnamefont {Leibfried}},\ and\
		\bibinfo {author} {\bibfnamefont {D.~J.}\ \bibnamefont {Wineland}},\
	}\bibfield  {title} {\bibinfo {title} {100-fold reduction of electric-field
			noise in an ion trap cleaned with in situ argon-ion-beam bombardment},\
	}\href {https://doi.org/10.1103/PhysRevLett.109.103001} {\bibfield  {journal}
		{\bibinfo  {journal} {Phys. Rev. Lett.}\ }\textbf {\bibinfo {volume} {109}},\
		\bibinfo {pages} {103001} (\bibinfo {year} {2012})}\BibitemShut {NoStop}%
	\bibitem [{\citenamefont {Kumph}\ \emph {et~al.}(2016)\citenamefont {Kumph},
		\citenamefont {Henkel}, \citenamefont {Rabl}, \citenamefont {Brownnutt},\
		and\ \citenamefont {Blatt}}]{kumph:2016}%
	\BibitemOpen
	\bibfield  {author} {\bibinfo {author} {\bibfnamefont {M.}~\bibnamefont
			{Kumph}}, \bibinfo {author} {\bibfnamefont {C.}~\bibnamefont {Henkel}},
		\bibinfo {author} {\bibfnamefont {P.}~\bibnamefont {Rabl}}, \bibinfo {author}
		{\bibfnamefont {M.}~\bibnamefont {Brownnutt}},\ and\ \bibinfo {author}
		{\bibfnamefont {R.}~\bibnamefont {Blatt}},\ }\bibfield  {title} {\bibinfo
		{title} {Electric-field noise above a thin dielectric layer on metal
			electrodes},\ }\href {https://doi.org/10.1088/1367-2630/18/2/023020}
	{\bibfield  {journal} {\bibinfo  {journal} {New Journal of Physics}\ }\textbf
		{\bibinfo {volume} {18}},\ \bibinfo {pages} {023020} (\bibinfo {year}
		{2016})}\BibitemShut {NoStop}%
	\bibitem [{\citenamefont {Brown}\ \emph {et~al.}(2016)\citenamefont {Brown},
		\citenamefont {Kim},\ and\ \citenamefont {Monroe}}]{brown:2016}%
	\BibitemOpen
	\bibfield  {author} {\bibinfo {author} {\bibfnamefont {K.~R.}\ \bibnamefont
			{Brown}}, \bibinfo {author} {\bibfnamefont {J.}~\bibnamefont {Kim}},\ and\
		\bibinfo {author} {\bibfnamefont {C.}~\bibnamefont {Monroe}},\ }\bibfield
	{title} {\bibinfo {title} {Co-designing a scalable quantum computer with
			trapped atomic ions},\ }\href {https://doi.org/10.1103/RevModPhys.87.1419}
	{\bibfield  {journal} {\bibinfo  {journal} {njp Quantum Inf}\ }\textbf
		{\bibinfo {volume} {2}},\ \bibinfo {pages} {16034} (\bibinfo {year}
		{2016})}\BibitemShut {NoStop}%
	\bibitem [{\citenamefont {Pino}\ \emph {et~al.}(2021)\citenamefont {Pino},
		\citenamefont {Dreiling}, \citenamefont {Figgatt}, \citenamefont {Gaebler},
		\citenamefont {Moses}, \citenamefont {H.Baldwin}, \citenamefont {Foss-Feig},
		\citenamefont {Hayes}, \citenamefont {Mayer}, \citenamefont {Ryan-Anderson},\
		and\ \citenamefont {Neyenhuis}}]{pino:2021}%
	\BibitemOpen
	\bibfield  {author} {\bibinfo {author} {\bibfnamefont {J.~M.}\ \bibnamefont
			{Pino}}, \bibinfo {author} {\bibfnamefont {J.~M.}\ \bibnamefont {Dreiling}},
		\bibinfo {author} {\bibfnamefont {C.}~\bibnamefont {Figgatt}}, \bibinfo
		{author} {\bibfnamefont {J.~P.}\ \bibnamefont {Gaebler}}, \bibinfo {author}
		{\bibfnamefont {S.~A.}\ \bibnamefont {Moses}}, \bibinfo {author}
		{\bibfnamefont {C.}~\bibnamefont {H.Baldwin}}, \bibinfo {author}
		{\bibfnamefont {M.}~\bibnamefont {Foss-Feig}}, \bibinfo {author}
		{\bibfnamefont {D.}~\bibnamefont {Hayes}}, \bibinfo {author} {\bibfnamefont
			{K.}~\bibnamefont {Mayer}}, \bibinfo {author} {\bibfnamefont
			{C.}~\bibnamefont {Ryan-Anderson}},\ and\ \bibinfo {author} {\bibfnamefont
			{B.}~\bibnamefont {Neyenhuis}},\ }\bibfield  {title} {\bibinfo {title}
		{Demonstration of the qccd trapped-ion quantum computer architecture},\
	}\href@noop {} {\bibfield  {journal} {\bibinfo  {journal} {Nature}\ }\textbf
		{\bibinfo {volume} {592}},\ \bibinfo {pages} {209} (\bibinfo {year}
		{2021})}\BibitemShut {NoStop}%
	\bibitem [{\citenamefont {Moehring}\ \emph {et~al.}(2011)\citenamefont
		{Moehring}, \citenamefont {Highstrete}, \citenamefont {Stick}, \citenamefont
		{Fortier}, \citenamefont {Haltli}, \citenamefont {Tigges},\ and\
		\citenamefont {Blain}}]{moehring:2011}%
	\BibitemOpen
	\bibfield  {author} {\bibinfo {author} {\bibfnamefont {D.~L.}\ \bibnamefont
			{Moehring}}, \bibinfo {author} {\bibfnamefont {C.}~\bibnamefont
			{Highstrete}}, \bibinfo {author} {\bibfnamefont {D.}~\bibnamefont {Stick}},
		\bibinfo {author} {\bibfnamefont {K.~M.}\ \bibnamefont {Fortier}}, \bibinfo
		{author} {\bibfnamefont {R.}~\bibnamefont {Haltli}}, \bibinfo {author}
		{\bibfnamefont {C.}~\bibnamefont {Tigges}},\ and\ \bibinfo {author}
		{\bibfnamefont {M.~G.}\ \bibnamefont {Blain}},\ }\bibfield  {title} {\bibinfo
		{title} {Design, fabrication and experimental demonstration of junction
			surface ion traps},\ }\href {https://doi.org/10.1088/1367-2630/13/7/075018}
	{\bibfield  {journal} {\bibinfo  {journal} {New Journal of Physics}\ }\textbf
		{\bibinfo {volume} {13}},\ \bibinfo {pages} {075018} (\bibinfo {year}
		{2011})}\BibitemShut {NoStop}%
	\bibitem [{\citenamefont {Johanning}\ \emph {et~al.}(2011)\citenamefont
		{Johanning}, \citenamefont {Braun}, \citenamefont {Eiteneuer}, \citenamefont
		{Paape}, \citenamefont {Balzer}, \citenamefont {Neuhauser},\ and\
		\citenamefont {Wunderlich}}]{johanning:2011}%
	\BibitemOpen
	\bibfield  {author} {\bibinfo {author} {\bibfnamefont {M.}~\bibnamefont
			{Johanning}}, \bibinfo {author} {\bibfnamefont {A.}~\bibnamefont {Braun}},
		\bibinfo {author} {\bibfnamefont {D.}~\bibnamefont {Eiteneuer}}, \bibinfo
		{author} {\bibfnamefont {C.}~\bibnamefont {Paape}}, \bibinfo {author}
		{\bibfnamefont {C.}~\bibnamefont {Balzer}}, \bibinfo {author} {\bibfnamefont
			{W.}~\bibnamefont {Neuhauser}},\ and\ \bibinfo {author} {\bibfnamefont
			{C.}~\bibnamefont {Wunderlich}},\ }\bibfield  {title} {\bibinfo {title}
		{Resonance-enhanced isotope-selective photoionization of ybi for ion trap
			loading},\ }\href {https://doi.org/10.1007/s00340-011-4502-7} {\bibfield
		{journal} {\bibinfo  {journal} {App. Phys. B}\ }\textbf {\bibinfo {volume}
			{103}},\ \bibinfo {pages} {327} (\bibinfo {year} {2011})}\BibitemShut
	{NoStop}%
	\bibitem [{\citenamefont {Roos}(2000)}]{roos:2000}%
	\BibitemOpen
	\bibfield  {author} {\bibinfo {author} {\bibfnamefont {C.~F.}\ \bibnamefont
			{Roos}},\ }\emph {\bibinfo {title} {Controlling the quantum state of trapped
			ions}},\ \href@noop {} {Ph.D. thesis},\ \bibinfo  {school} {University of
		Innsbruck} (\bibinfo {year} {2000})\BibitemShut {NoStop}%
	\bibitem [{\citenamefont {Noel}\ \emph {et~al.}(2019)\citenamefont {Noel},
		\citenamefont {Berlin-Udi}, \citenamefont {Matthiesen}, \citenamefont {Yu},
		\citenamefont {Zhou}, \citenamefont {Lordi},\ and\ \citenamefont
		{Haffner}}]{noel:2019}%
	\BibitemOpen
	\bibfield  {author} {\bibinfo {author} {\bibfnamefont {C.}~\bibnamefont
			{Noel}}, \bibinfo {author} {\bibfnamefont {M.}~\bibnamefont {Berlin-Udi}},
		\bibinfo {author} {\bibfnamefont {C.}~\bibnamefont {Matthiesen}}, \bibinfo
		{author} {\bibfnamefont {J.}~\bibnamefont {Yu}}, \bibinfo {author}
		{\bibfnamefont {Y.}~\bibnamefont {Zhou}}, \bibinfo {author} {\bibfnamefont
			{V.}~\bibnamefont {Lordi}},\ and\ \bibinfo {author} {\bibfnamefont
			{H.}~\bibnamefont {Haffner}},\ }\bibfield  {title} {\bibinfo {title}
		{Electric-field noise from thermally activated fluctuators in a surface ion
			trap},\ }\href@noop {} {\bibfield  {journal} {\bibinfo  {journal} {Phys. Rev.
				A}\ }\textbf {\bibinfo {volume} {99}},\ \bibinfo {pages} {063427} (\bibinfo
		{year} {2019})}\BibitemShut {NoStop}%
	\bibitem [{\citenamefont {An}\ \emph {et~al.}(2019)\citenamefont {An},
		\citenamefont {Matthiesen}, \citenamefont {Urban},\ and\ \citenamefont
		{Haffner}}]{an:2019}%
	\BibitemOpen
	\bibfield  {author} {\bibinfo {author} {\bibfnamefont {D.}~\bibnamefont
			{An}}, \bibinfo {author} {\bibfnamefont {C.}~\bibnamefont {Matthiesen}},
		\bibinfo {author} {\bibfnamefont {E.}~\bibnamefont {Urban}},\ and\ \bibinfo
		{author} {\bibfnamefont {H.}~\bibnamefont {Haffner}},\ }\bibfield  {title}
	{\bibinfo {title} {Distance scaling and polarization of electric-field noise
			in a surface ion trap},\ }\href {https://doi.org/10.1103/PhysRevA.100.063405}
	{\bibfield  {journal} {\bibinfo  {journal} {Phys. Rev. A}\ }\textbf {\bibinfo
			{volume} {100}},\ \bibinfo {pages} {063405} (\bibinfo {year}
		{2019})}\BibitemShut {NoStop}%
	\bibitem [{\citenamefont {Harlander}\ \emph {et~al.}(2012)\citenamefont
		{Harlander}, \citenamefont {Brownnutt}, \citenamefont {Hänsel},\ and\
		\citenamefont {Blatt}}]{harlander:2010}%
	\BibitemOpen
	\bibfield  {author} {\bibinfo {author} {\bibfnamefont {M.}~\bibnamefont
			{Harlander}}, \bibinfo {author} {\bibfnamefont {M.}~\bibnamefont
			{Brownnutt}}, \bibinfo {author} {\bibfnamefont {W.}~\bibnamefont {Hänsel}},\
		and\ \bibinfo {author} {\bibfnamefont {R.}~\bibnamefont {Blatt}},\ }\bibfield
	{title} {\bibinfo {title} {Trapped-ion probing of light induced charging
			effects on dielectrics},\ }\href
	{https://doi.org/10.1088/1367-2630/12/9/093035} {\bibfield  {journal}
		{\bibinfo  {journal} {New J. Phys.}\ }\textbf {\bibinfo {volume} {12}},\
		\bibinfo {pages} {093035} (\bibinfo {year} {2012})}\BibitemShut {NoStop}%
\end{thebibliography}

\providecommand{\noopsort}[1]{}\providecommand{\singleletter}[1]{#1}%

\end{document}